\documentclass{article}
\usepackage{float}
\usepackage{url}
\usepackage{pdflscape}
\usepackage{graphicx}
\usepackage{amsmath,amsfonts,amssymb}
\usepackage{subcaption}
\usepackage{epsfig,times,latexsym,enumerate,lscape}
\usepackage{textcomp}

\begin{document}

\title{\large\bf{AESHA3: Efficient and Secure Sub-Key Generation for AES Using SHA-3}}
\author{Ankush Soni\footnote{Dept. of CSIS, BITS Pilani K.K. Birla Goa Campus, Goa, India, Email: p20180413@goa.bits-pilani.ac.in} \hspace{0.15mm} Sanjay K. Sahay\footnote{Dept. of CSIS, BITS Pilani K.K. Birla Goa Campus, Goa, India, Email: ssahay@goa.bits-pilani.ac.in} \hspace{0.15mm} Parit Mehta\footnote{Dept. of CSIS, BITS Pilani K.K. Birla Goa Campus, Goa, India, Email: h20210036g@alumni.bits-pilani.ac.in}}
\date{}
\maketitle

\begin{abstract}
Advanced Encryption Standard (AES) is one of the most widely used symmetric cipher for the confidentiality of data. Also it is used for other security services, viz. integrity, authentication and key establishment. However, recently, authors have shown some weakness in the generation of sub-keys in AES, e.g. bit leakage attack, etc. Also, AES sub-keys are generated sequentially, which is an overhead, especially for resource-constrained devices. Therefore, we propose and investigate a novel encryption AESHA3, which uses sub-keys generated by Secure Hash Algorithm-3 (SHA3). The output of SHA3 is one-way and highly non-linear, and random. The experimental analysis shows that the average time taken for generating the sub-keys to be used for encrypting the data using our approach i.e. AESHA3 is $\sim$ 1300 times faster than the sub-key generated by the standard AES. Accordingly, we find that AESHA3 will be very relevant not only in terms of security but also it will save the resources in IoT devices. We investigated AESHA3 in Intel Core i7, 6th Generation processor and Raspberry Pi 4B and found that up to two MB data encryption is very significant, and lesser the data size, more the resource saving compared to AES.

\it{Advanced Encryption Standard, Sub-keys, Secure Hash Algorithm-3, IoT Devices}
\end{abstract}

\section{Introduction}
Advanced Encryption Standard (AES) is the most widely used symmetric key cipher, basically for data confidentiality. However it can provide other security services viz. integrity, authentication and key establishment in several real-life applications, including financial services, data centres, web security, etc. The AES has three variant based on the key size i.e. AES-128, AES-192 and AES-256. The design of AES is cryptographically strong \cite{aes}, and to date, no attack better than brute force has been found. Additionally, because of the 256-bit key variant, it might even cope with the offing quantum computations for the next two decades. However, the sub-key generation process of AES from the master key is arguably frail \cite{May2002}. Also, with the advent of smart devices and the Internet of Things (IoT) over the last few years, there is a high demand for an efficient and secure algorithm for these resource and energy-constrained devices. Therefore in this paper we propose a novel efficient and secure sub-key generation for AES using Secure Hash Algorithm-3 (SHA-3) for security services, and named AESHA3. We use the same three fundamental layers of AES (Key Addition, Shift Rows and Mix Columns) approved by the National Institute of Standards and Technology (NIST) \cite{aes}. However the sub-keys are generated from SHA3, which is highly random, non-linear and one-way (i.e. the key cannot be generated in the reverse order) \cite{sha3} i.e., it has all the properties that the NIST-approved AES sub-keys posses. Hence, we have used it to generate the sub-keys of all rounds of AES for the encryption of the data.

The remainder of the paper is divided as follows: Section II briefly discusses the related work. Section III gives the brief description of AES and SHA3. While section IV discusses the problem overview and our approach for sub-key generation for AESHA3. Section V and VI describes the experimental setup and analysis of our results respectively. Finally, the section VII contains the conclusion of the paper.

\section{Related Work}
\begin{sloppypar} % Added to address overfull \hbox
According to Kerckhoff's Principles \cite{Petitcolas2011}, a cipher or cryptosystem should be secure even if the attacker knows all details about the system, except the secret key i.e. the system should be secure even if the attacker knows the encryption and decryption processes. Therefore, the keys used for any cipher are the most important aspect of achieving the desired security level. In symmetric cipher, generally sub-keys are generated from the selected key so that even if the selected key is weak, the sub-key generation process makes the sub-keys highly random so that security of the cipher entirely lies on the sub-keys. Therefore, time-to-time researchers investigated one of the widely used AES sub-keys generation processes. The first weakness in the AES key schedule was discovered by Lauren and Matt in 2002 \cite{May2002}. According to them, the key schedule of the AES is vulnerable to bit leakage i.e. even with partial knowledge of the previous sub-key, all the other sub-keys can be generated. They proposed a rectified key schedule in which every subkey depends on every bit of the original key. Their proposed key schedule secures the keys from bit leakage. However, the computational resources required for generating the sub-keys remain more or less the same as the original AES sub-key generation processes. Later in 2008, Bahrak et al. \cite{Bahrak2008} proposed a differential attack which exploits differences at the intermediate state of the AES algorithm and time-memory trade off. They have shown that the best differential attack requires $2^{115.5}$ chosen plain-texts, $2^{109}$ bytes of memory and seven rounds only to attack AES-128. However, it is almost impossible to get $2^{115.5}$ of plain-texts.

In 2010, Biryukov et al. \cite{Biryukov1999} show that AES-192 and AES-256 security level can be 176 and 199 bits respectively. However, still, it is completely impractical to find the key. The reason for their reduced security level may be due to generating 1.5 and 2 sub-keys with the standard AES key schedule. Therefore, their attack is not applicable to AES-128 because one sub-key is generated in each iteration in AES-128. Also, their attack depends on how the keys are related.

Understanding the importance of light weight cipher for IoT devices, Bogdanov et al. \cite{Bogdanov2014} in 2014 proposed a light weight encryption scheme which uses AES-128 key schedule. The cipher is an online, single-pass authenticated encryption algorithm that supports optional associated data, and its security relies on nonces. Later in 2019 Alasaad et al. \cite{Alasaad2019} claimed that as the S-box used in AES is a static and fixed matrix, therefore, a backdoor can be built into the cipher to exploit the AES. Hence, they proposed a simple key dependent S-box scheme which generates a dynamic S-box for each round of encryption. Their approach uses some bits of the primary key to directly manipulate the standard S-box in such a way that its content is changed but its cryptographic properties are preserved. They have shown that their proposed method strengthens the cipher against certain attacks at the expense of a relatively modest one-time computational procedure during the set-up phase.

In 2020, Leurent et al. \cite{Leurent2021}, proposed a key schedule in which all the sub-keys of the rounds are generated independently of each other, unlike the original AES key schedule, where the sub-keys are generated from a single master key. Their proposed approach security level remains the same as that of the original AES. Also, the number of steps for generating the sub-keys was exactly the same. Later in 2021, G. Leurent et al. \cite{Leurent2021} proposed a modified AES key schedule in which all the sub-keys of the rounds are generated independently, unlike the original AES key schedule, where the sub-keys are generated sequentially from the master key. Their basic idea is to parallelise the computation of the key schedule. They generate an equivalent representation of the AES key schedule using invariant subspace attacks. However, the number of computations for the variants of AES are the same as the original AES.

Recently, Sawka et al. \cite{Sawka2022} proposed a scheme to create key expansion algorithms for modern block ciphers. Their approach uses the sponge construction for creating the key expansion algorithm. Further, their method uses the key expansion algorithm on a specifically designed block cipher, IJON. As their approach was specific to the custom block cipher, its scalability is not guaranteed.
\end{sloppypar}

\section{Brief Description of Advanced Encryption Standard and Secure Hash Algorithm-3}

\subsection{Advanced Encryption Standard}
The first approved block cipher was Data Encryption Standard (DES) \cite{DES}, which is still secure as per its design, i.e. no attack better than brute force attack has been found yet. However, due to its small key size (56 bits), DES is no longer recommended; instead, one can use 3DES, which is three times slower than DES and provides 112-bit security only. Therefore, to have an efficient and more secure symmetric cipher, NIST approved AES in 2002 \cite{aes}. In general, symmetric block cipher processes $n$ bits of plain text and $m$ bits of the key to generate $n$ bits of random cipher text. The security of the modern symmetric cipher is achieved by applying two characteristics, i.e. confusion and diffusion, introduced by Claude Shannon \cite{Shannon} in multiple rounds (in AES, the number of rounds depends on the key size to produce a completely random ciphertext). The confusion property obscures the relationship between the key and cipher text, while diffusion obscures the relationship between plain text and cipher text to resist cryptanalysis. AES transforms the input data into three layers, i.e. substitution layer for confusion property, diffusion layer for diffusion property and in key addition layer, data state is XORed with the generated non-linear sub-keys to make the cipher random.

The AES operates on bytes rather than bits, and its input (plaintext) of 128 bits is represented as 16 bytes, arranged in a 4 x 4 matrix. This plaintext of the matrix is known as the initial state and is modified as the algorithm progresses. AES comes in three different variants viz. 128, 192, and 256 input key bits. Each of these is arranged in a 4 x 4/6/8 matrix, each column representing a word of 4 bytes. However, in encryption processes, only 128 bits are used to XOR the data state irrespective of the main key size. Therefore, the number of rounds of encryption processes increases with key size and are 10, 12 and 14 rounds for 128, 192 and 256-bit key sizes, respectively. Table \ref{table:round_details} shows the number of rounds, number of sub-keys and total number of bits generated to get the ciphertext. The number of sub-keys is one more than the number of rounds because before the start of the encryption processes, the original plaintext is initially XORed with the first subkey (master-key), and this XORing of plaintext before the start of the actual encryption processes is known as key whitening.

\begin{table}
\begin{center}
\caption{Number of rounds and sub-keys in the variants of AES}
\label{table:round_details}
\begin{tabular}{|c|c|c|c|}
\hline
\textbf{AES Version}           & \textbf{AES-128} & \textbf{AES-192} & \textbf{AES-256} \\ \hline
Number of bits in the sub-key & 128              & 192              & 256              \\ \hline
Number of rounds               & 10               & 12               & 14               \\ \hline
Number of sub-keys            & 11               & 13               & 15               \\ \hline
Total number of bits required & 1408             & 1664             & 1920             \\ \hline
\end{tabular}
\end{center}
\end{table}

\subsection{Secure Hash Algorithm-3}
A cryptographic hash function (CHF) is a one-way function which takes arbitrary size input and provides a highly compressed fixed-size output. It is key-less but plays a very important role in digital security, viz., checks the integrity of the data because it is deterministic and highly sensitive (on average, when a single bit in the data is changed, approximately half of the bits in the hash output will be modified), and it is computationally impossible to find another message for the known hash value. Therefore, it is also known as checksum, message digest and fingerprint. The security level of a hash function is determined by computationally in-feasibility to find two different inputs that generate the same hash, i.e. collision resistance, and to find the original input from the hash value, i.e. preimage-resistance. Generally, if $n$ is the size of the hash value, then a good hash function shall have $n/2$ bits security level.

\begin{figure}[H]
\centering
\includegraphics[width=\linewidth]{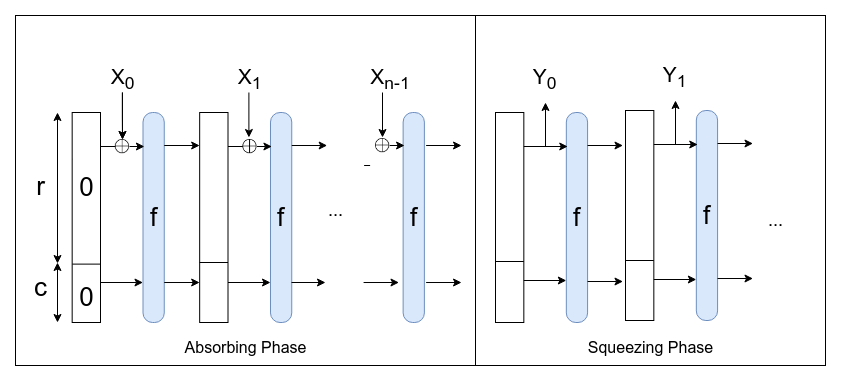}
\caption{A schematic of SHA-3}
\label{fig:sponge}
\end{figure}

The first CHF was designed by Rabin in the late 70s. He proposed a 64-bit hash function using DES block cipher \cite{Preneel2010}. Later, in the 80s, more hash functions were proposed, and in 90s, MD5 and SHA-1 were used in several applications \cite{sha3}. However, in 2004, it was shown that finding the collision in Message Digest-5 (MD5) is easy, and the security level of SHA-1 is significantly less than the standard required security. Hence SHA-2 series (SHA-256, SHA-384, SHA-512) has been approved, and to date, no collision has been found in SHA-2. However, understanding the significance of CHF for security services, in 2012, NIST approved SHA-3 \cite{sha3} to meet out the digital security requirements. The design of SHA-3 is very different from earlier hash functions. A schematic of the SHA-3 is shown in figure \ref{fig:sponge}, which is based on sponge construction. It has two phases: in the first phase, the message is absorbed into the sponge, and in the second phase, the result is squeezed. In the absorbing phase, message blocks ($x_i$) are XORed into a subset of the state and the output blocks are read from the same subset of the state and alternated with the state transformation function ($f$). The size of the part of the message that is written and read is known as the \emph{rate} (r), and the size of the part that is untouched by input/output is known as the \emph{capacity} (c). The capacity determines the security of the SHA-3 and is c/2, and the security level does not changes even one take more than the double the bit of the desired security level from the output of the SHA-3. The message transformation is done by the $f$-function in five steps called $\theta, \; \rho, \; \pi, \; \chi$ and $\iota$ in which computation of the parity, bit-wise rotation, permutation of 25 words, bit-wise combination along rows and exclusive-OR are done respectively. To get the hash, after the absorption is completed, the function $f$ of the state is taken until the required length is obtained, i.e. Hash(x) = $Y_o || Y_1 || ...$

\begin{figure}[ht]
\begin{subfigure}{.5\textwidth}
  \centering
  \includegraphics[width=.8\linewidth]{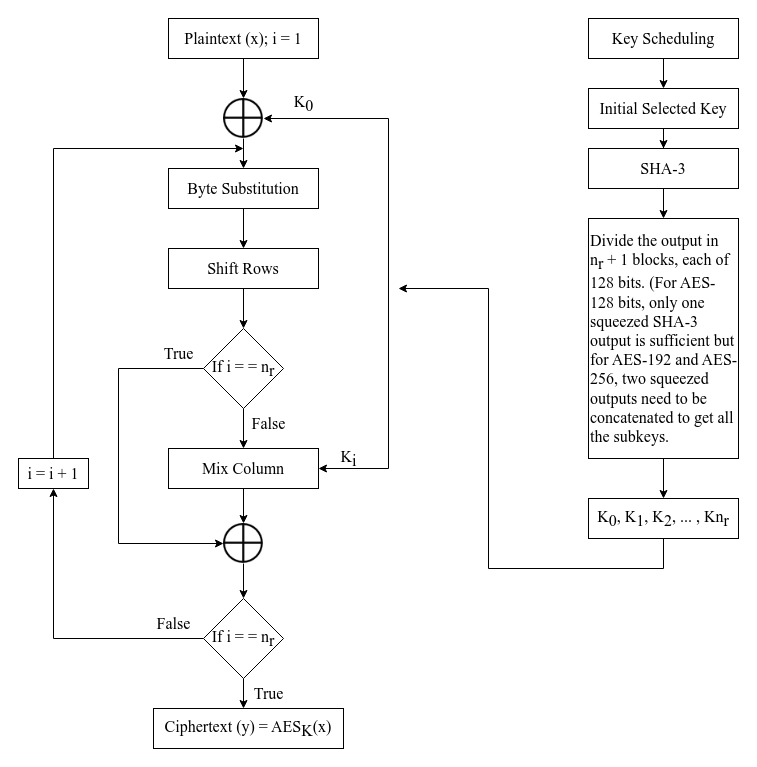}
  \caption{AES Encryption}
  \label{aes_encr}
\end{subfigure}
\begin{subfigure}{.5\textwidth}
  \centering
  \includegraphics[width=.8\linewidth]{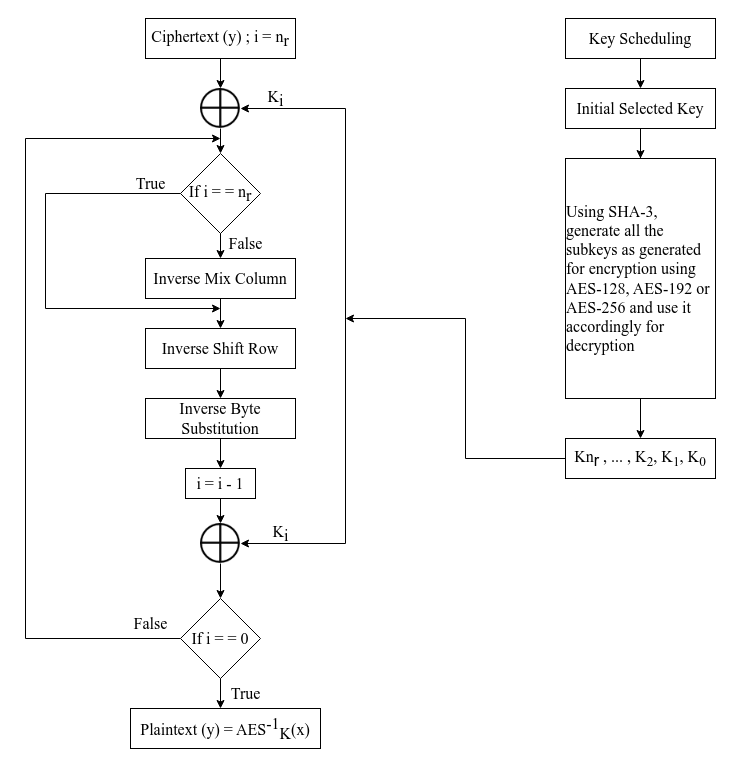}
  \caption{AES Decryption}
  \label{aes_decr}
\end{subfigure}
\caption{Flowchart for AES}
\end{figure}

\section{Problem Overview and Proposed Approach}

\subsection{Problem Overview}
In AES, the round sub-keys are generated sequentially from the master key using a non-linear $g$-function, which uses the same S-box that is used in the byte substitution layer \cite{aes}. Depending on key size, the number of words (32 bits) are generated. This sequential generation of keys creates additional computational overhead, resulting in more resource utilisation, especially in resource-constrained IoT devices. Also, the authors \cite{aes, Leurent2021, DeLosReyes2019, Nikolic2011} pointed out the weakness in the key generation process of the AES. Therefore, we investigate SHA-3 for efficient and secure sub-key generation processes for data encryption and decryption.

\subsection{Proposed Approach}
We propose and investigate a novel and efficient approach for generating the sub-keys using SHA-3. The output of the squeezed phase in SHA-3 provides a non-linear one-way 1600-bit output and can be used as sub-keys in the encryption of data in AES. The required number of bits for encrypting data in different variant of AES is shown in Table \ref{table:round_details}. For AES-128, just first output of the squeezing phase is sufficient to have all the 11 sub-keys. However, in case of AES-192 and AES-256, one more iteration of the squeezing phase had to run to get all the sub-keys to encrypt the data, i.e. one can have 3200 bits of non-linear bits to make 13 or 15 sub-keys, i.e.
\begin{center}
Hash(x) = $Y_o || Y_1$
\end{center}
where $x$ denotes the input string, $Y_o$ and $Y_1$ are the output of SHA-3 squeezing phase, which is of 1600 bits. The processes of squeezing phase in such that from $Y_1$ it is computationally infeasible to find $Y_o$ and taking any sequence of 128 bits from Hash(x) does not effect the security of the sub-keys.
Figures \ref{aes_encr} and \ref{aes_decr} show the flowchart of AES encryption and decryption processes of AESHA3.

\section{Experimental Setup}
To investigate the proposed novel approach, we implemented it on Intel Core i7 6th generation, 12 GB RAM, in Ubuntu 22.04 LTS OS and Raspberry Pi 4, 8 GB RAM with Raspbian OS using Python 3.10. First, we run the SHA-3 10,000 times with random inputs strings and found the average time taken for generating the sub-keys to be used for encrypting the data using the AES three layer is $\sim$ 1300 times faster then sub-key generated by he standard AES. The detail results are shown in table \ref{table:Results}. Then we used Electronic Code Book modes of operations i.e. encrypting data of 128 bits or more independently to find how much time can be saved for data encryption for IoT devices, which are generally resource constrained.

\begin{figure}
	\centering
	\begin{minipage}[h]{\textwidth}
	    \centering
    	\includegraphics[width=\linewidth]{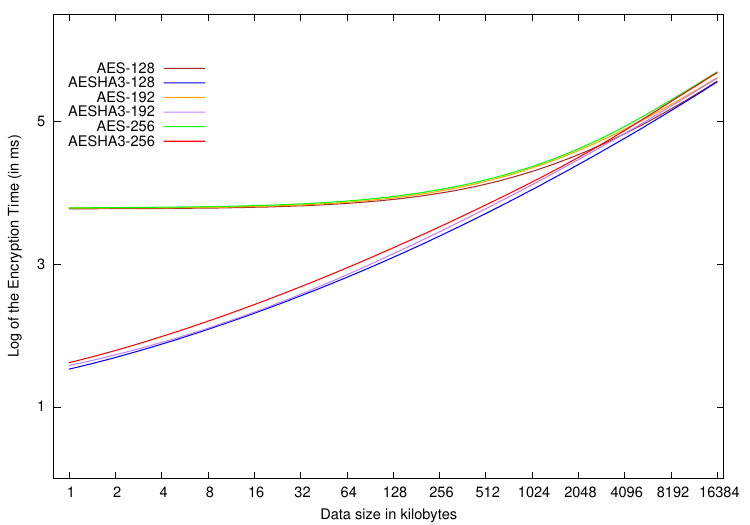}
    	\caption{Comparison of the time taken to encrypt the data by AES and AESHA3 by intel core i7, 6th generation processor.}
    	\label{fig:total_rpi4}
	\end{minipage}
	\hfill
	\begin{minipage}[h]{\textwidth}
    	\centering
    	\includegraphics[width=0.9\linewidth]{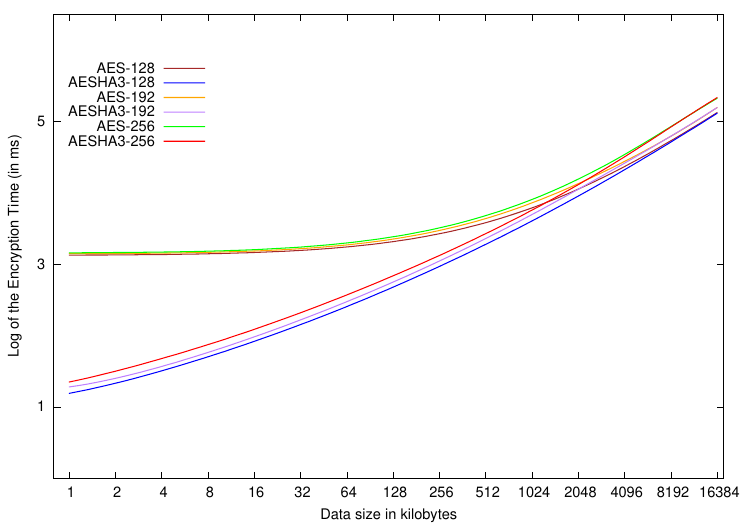}
    	\caption{Comparison of the time taken to encrypt the data by AES and AESHA3 by Raspberry Pi 4B.}
    	\label{fig:total_i7}
    	\end{minipage}
\end{figure}

\begin{figure}
    \begin{minipage}[h]{\textwidth}
	    \centering
    	\includegraphics[width=\linewidth]{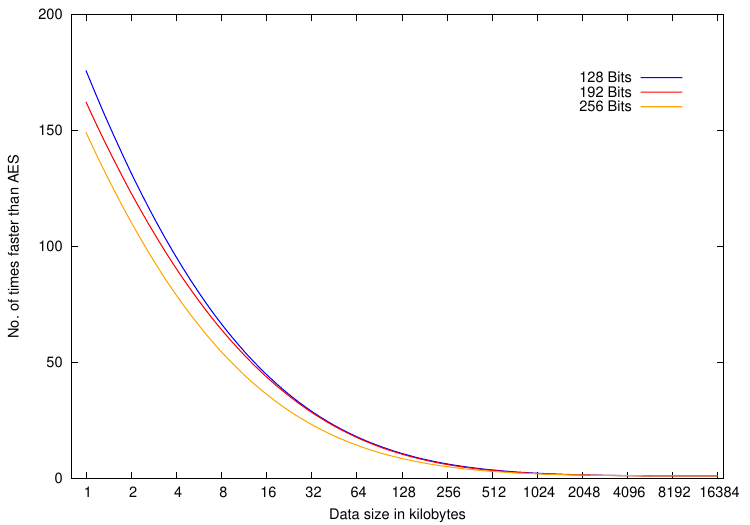}
    	\caption{Number of times AESHA3 is faster than AES to encrypt data by Raspberry Pi 4B.}
    	\label{fig:efficiency_rpi}
	\end{minipage}
	\hfill
	\begin{minipage}[h]{\textwidth}
    	\centering
    	\includegraphics[width=0.9\linewidth]{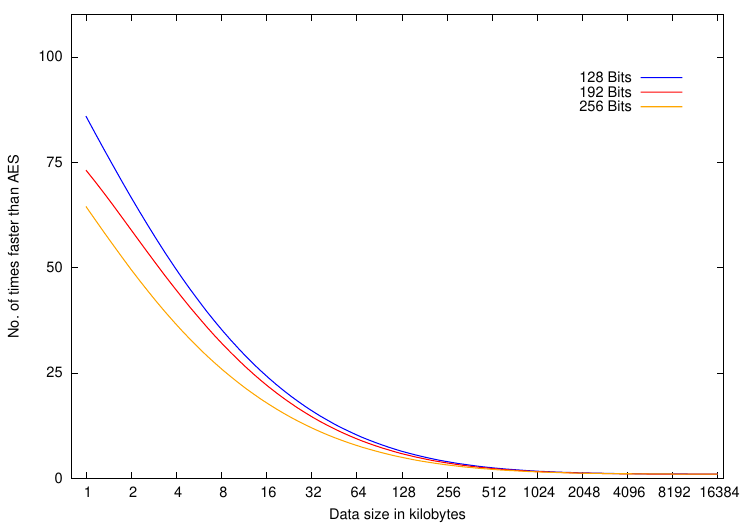}
    	\caption{Number of times AESHA3 faster then AES to encrypt data by intel core i7, 6th generation processor.}
    	\label{fig:efficiency_i7}
    	\end{minipage}
\end{figure}

\begin{table*}[h]
\centering
\caption{Time taken in milliseconds for the sub-key generation by AES and AESHA3}
\resizebox{\textwidth}{!}{
\begin{tabular}{|c|c|c|c|c|c|c|c|}
\hline
\label{table:Results}
\textbf{\begin{tabular}[c]{@{}c@{}}System\\ Specifications\end{tabular}}                     & \textbf{\begin{tabular}[c]{@{}c@{}}Operating\\ System\end{tabular}} & \textbf{\begin{tabular}[c]{@{}c@{}} AES-128 \end{tabular}} & \textbf{\begin{tabular}[c]{@{}c@{}}AESHA3-128 \end{tabular}} & \textbf{\begin{tabular}[c]{@{}c@{}} AES-192 \end{tabular}} & \textbf{\begin{tabular}[c]{@{}c@{}} AESHA3-192 \end{tabular}} & \textbf{\begin{tabular}[c]{@{}c@{}} AES-256 \end{tabular}} & \textbf{\begin{tabular}[c]{@{}c@{}}AESHA3-256 \end{tabular}} \\
\hline
\begin{tabular}[c]{@{}c@{}}Intel Core i7\\ 6th generation CPU\end{tabular} & \begin{tabular}[c]{@{}c@{}}Ubuntu 22.04\\ LTS\end{tabular}          & 1334.31                                                                            & 1.21                                                                               & 1403.26                                                                            & 1.231                                                                              & 1442.18                                                                            & 1.237 \\
\hline
\begin{tabular}[c]{@{}c@{}}Raspberry Pi 4B\\ 8GB RAM\end{tabular}                            & Raspbian OS                                                         & 5996.36                                                                            & 4.55                                                                               & 6084.61                                                                            & 4.74                                                                               & 6205.45                                                                            & 4.69                                                                               \\ \hline
\end{tabular}}
\end{table*}

\begin{table}[h]
\scriptsize
\caption{Time taken and efficiency to encrypt the data up to 16 MB using AES and AESHA3 in Intel Core i7 6th Generation processor.}
\centering
\resizebox{\textwidth}{!}{
\renewcommand{\arraystretch}{1.5}
\begin{tabular}{|c|c|c|c|c|c|c|c|c|c|}
\hline
\label{table:i7_results}
\textbf{File Size} & \textbf{\begin{tabular}[c]{@{}c@{}}Total Time\\ AES-128 \\(ms)\end{tabular}} & \textbf{\begin{tabular}[c]{@{}c@{}}Total Time\\ AESHA3-128 \\(ms)\end{tabular}} & \textbf{\begin{tabular}[c]{@{}c@{}}Efficiency of \\ AESHA3-128 \\ (X) \end{tabular}} & \textbf{\begin{tabular}[c]{@{}c@{}}Total Time \\ AES-192 \\(ms)\end{tabular}} & \textbf{\begin{tabular}[c]{@{}c@{}}Total Time\\ AESHA3-192 \\(ms)\end{tabular}} & \textbf{\begin{tabular}[c]{@{}c@{}}Efficiency of \\ AESHA3-192 \\ (X)\end{tabular}} & \textbf{\begin{tabular}[c]{@{}c@{}}Total Time \\ AES-256 \\(ms)\end{tabular}} & \textbf{\begin{tabular}[c]{@{}c@{}}Total Time\\ AESHA3-256 \\(ms)\end{tabular}} & \textbf{\begin{tabular}[c]{@{}c@{}}Efficiency of \\ AESHA3-256 \\ (X)\end{tabular}} \\ \hline
\textbf{1 KB} & 1347.24 & 15.68 & 85.92 & 1417.22 & 19.38 & 73.12 & 1464.15 & 22.70 & 64.5 \\ \hline
\textbf{2 KB} & 1355.44 & 23.69 & 57.21 & 1424.68 & 26.26 & 54.25 & 1476.69 & 34.88 & 42.33 \\ \hline
\textbf{4 KB} & 1373.93 & 42.29 & 32.49 & 1448.18 & 49.99 & 28.97 & 1509.12 & 68.89 & 21.91 \\ \hline
\textbf{8 KB} & 1413.34 & 81.60 & 17.32 & 1491.14 & 92.22 & 16.17 & 1548.31 & 106.64 & 14.52 \\ \hline
\textbf{16 KB} & 1473.12 & 141.49 & 10.41 & 1566.51 & 168.31 & 9.31 & 1665.26 & 222.89 & 7.47 \\ \hline
\textbf{32 KB} & 1604.57 & 272.77 & 5.88 & 1729.70 & 330.97 & 5.22 & 1861.00 & 418.86 & 4.44 \\ \hline
\textbf{64 KB} & 1881.59 & 550.00 & 3.42 & 2044.98 & 646.93 & 3.16 & 2225.71 & 783.58 & 2.84 \\ \hline
\textbf{128 KB} & 2402.95 & 1071.30 & 2.24 & 2656.92 & 1257.83 & 2.43 & 2903.74 & 1462.38 & 1.98 \\ \hline
\textbf{256 KB} & 3388.58 & 2056.83 & 1.65 & 3892.48 & 2493.95 & 1.56 & 4358.85 & 2917.59 & 1.49 \\ \hline
\textbf{512 KB} & 5461.41 & 4129.83 & 1.32 & 6398.85 & 4999.23 & 1.28 & 7278.61 & 5836.77 & 1.25 \\ \hline
\textbf{1 MB} & 9649.51 & 8317.80 & 1.16 & 12831.64 & 11432.31 & 1.12 & 13094.25 & 11653.24 & 1.12 \\ \hline
\textbf{2 MB} & 17779.87 & 16448.04 & 1.08 & 22338.69 & 20939.92 & 1.06 & 24701.81 & 23259.94 & 1.062 \\ \hline
\textbf{4 MB} & 35315.09 & 33984.14 & 1.04 & 40327.12 & 38924.27 & 1.03 & 57461.65 & 56015.52 & 1.025 \\ \hline
\textbf{8 MB} & 67805.52 & 66472.24 & 1.02 & 80735.64 & 79330.45 & 1.02 & 114207.08 & 112728.99 & 1.013 \\ \hline
\textbf{16 MB} & 134256.81 & 132920.93 & 1.01 & 160978.36 & 159569.76 & 1.01 & 219933.49 & 218440.19 & 1.006 \\ \hline
\end{tabular}}
\end{table}

\begin{table}[h]
\caption{Time taken and efficiency to encrypt the data up to 16 MB using AES and AESHA3 in Raspberry Pi 4B.}
\scriptsize
\centering
\resizebox{\textwidth}{!}{
\renewcommand{\arraystretch}{1.5}
\begin{tabular}{|c|c|c|c|c|c|c|c|c|c|}
\hline
\label{table:raspi4_results}
\textbf{File Size} & \textbf{\begin{tabular}[c]{@{}c@{}}Total Time \\ AES-128 \\ (ms)\end{tabular}} & \textbf{\begin{tabular}[c]{@{}c@{}}Total Time\\ AESHA3-128 \\ (ms)\end{tabular}} & \textbf{\begin{tabular}[c]{@{}c@{}}Efficiency of\\ AESHA3-128 \\(X) \end{tabular}} & \textbf{\begin{tabular}[c]{@{}c@{}}Total Time\\ AES-192 \\ (ms)\end{tabular}} & \textbf{\begin{tabular}[c]{@{}c@{}}Total Time\\ AESHA3-192 \\(ms)\end{tabular}} & \textbf{\begin{tabular}[c]{@{}c@{}}Efficiency of\\ AESHA3-192 \\ (X) \end{tabular}} & \textbf{\begin{tabular}[c]{@{}c@{}}Total Time\\ AES-256 \\ (ms)\end{tabular}} & \textbf{\begin{tabular}[c]{@{}c@{}}Total Time\\ AESHA3-256 \\ (ms) \end{tabular}} & \textbf{\begin{tabular}[c]{@{}c@{}}Efficiency of\\ AESHA3-256 \\(X)\end{tabular}} \\ \hline
\textbf{1 KB} & 6,021.61 & 34.22 & 175.97 & 6,109.35 & 37.62 & 162.40 & 6,230.05 & 41.73 & 149.29 \\ \hline
\textbf{2 KB} & 6,044.45 & 56.67 & 106.66 & 6,131.04 & 60.83 & 100.79 & 6,258.88 & 71.06 & 88.08 \\ \hline
\textbf{4 KB} & 6,088.72 & 101.13 & 60.21 & 6,174.92 & 103.58 & 59.61 & 6,319.90 & 131.98 & 47.89 \\ \hline
\textbf{8 KB} & 6,185.17 & 197.83 & 31.27 & 6,263.59 & 190.93 & 32.81 & 6,445.41 & 256.86 & 25.09 \\ \hline
\textbf{16 KB} & 6,351.80 & 364.67 & 17.42 & 6,445.58 & 373.80 & 17.24 & 6,685.70 & 497.24 & 13.45 \\ \hline
\textbf{32 KB} & 6,700.51 & 713.00 & 9.40 & 6,853.98 & 782.73 & 8.76 & 7,158.61 & 970.84 & 7.37 \\ \hline
\textbf{64 KB} & 7,393.14 & 1,406.06 & 5.26 & 7,722.23 & 1,651.22 & 4.68 & 8,126.02 & 1,938.22 & 4.19 \\ \hline
\textbf{128 KB} & 8,790.51 & 2,803.58 & 3.14 & 9,437.66 & 3,366.87 & 2.80 & 10,078.41 & 3,891.07 & 2.59 \\ \hline
\textbf{256 KB} & 11,656.09 & 5,668.24 & 2.06 & 12,743.25 & 6,675.07 & 1.91 & 13,874.86 & 7,689.83 & 1.80 \\ \hline
\textbf{512 KB} & 17,375.71 & 11,388.39 & 1.53 & 19,583.15 & 13,510.26 & 1.45 & 21,684.56 & 15,499.15 & 1.40 \\ \hline
\textbf{1 MB} & 28,641.21 & 22,653.57 & 1.26 & 32,821.73 & 26,749.37 & 1.23 & 28,572.75 & 22,384.93 & 1.28 \\ \hline
\textbf{2 MB} & 51,266.41 & 45,279.09 & 1.13 & 59,780.93 & 53,657.67 & 1.11 & 67,857.84 & 61,664.73 & 1.10 \\ \hline
\textbf{4 MB} & 98,149.74 & 92,156.76 & 1.07 & 111,105.16 & 104,965.89 & 1.06 & 129,221.49 & 123,033.96 & 1.05 \\ \hline
\textbf{8 MB} & 187,024.09 & 180,998.08 & 1.03 & 218,761.47 & 212,593.99 & 1.03 & 255,063.21 & 248,797.29 & 1.03 \\ \hline
\textbf{16 MB} & 368,645.63 & 362,567.91 & 1.02 & 415,755.80 & 409,473.02 & 1.02 & 499,292.89 & 493,065.41 & 1.01 \\ \hline
\end{tabular}}
\end{table}

\section{Results Analysis}
As we find that sub-key generated by SHA-3 takes $\sim$ 1300 less time than the sub-key generated by AES. Therefore for IoT devices we investigated that how much it will be efficient if data is encrypted by AESHA3. For the purpose first we implemented AESHA3 and AES to encrypt data in intel core i7, 6th generation processor and then we implemented in Raspberry Pi 4B because it has limited computational power and can be considered as IoT device. We encrypted the data in both operating systems starting from 1 KB and doubling the data till 16 MB. We find that up to 2 MB data encryption it is significant, and lesser the data size more the resource saving (table \ref{table:i7_results} and \ref{table:raspi4_results}). This is basically due to fact that sub-key generation is one time job. The figure \ref{fig:total_rpi4} and \ref{fig:total_i7} shows the time taken to encrypt the data with AESHA3 and AES upto 16 MB, and figure \ref{fig:efficiency_rpi} and \ref{fig:efficiency_i7} shows the number of times AESHA3 is faster then AES in intel core i7, 6th generation processor and Rasberry Pi 4B respectively.

\pagebreak

\section{Conclusion}
In symmetric cipher, AES is the most widely used cipher for the confidentiality of the data and is also used for the security services viz. integrity, authentication and key establishment. According to Kerckhoff’s Principles, a cipher or cryptosystem should be secure even if the attacker knows all details about the system, except the secret key, i.e. the system should be secure even if the attacker knows the encryption/decryption algorithms. However, recently authors have shown some weakness in the generation of sub-keys in AES, hence a threat to the AES cipher. Therefore, we proposed and investigated a novel encryption AESHA3, which uses sub-key generated by SHA3, which provides a one way and highly non-linear output i.e. the overall security will not change. However, our analysis shows that the average time taken for generating the sub-keys to be used for encrypting the data using the three layers of AES is $\sim$ 1300 times faster than sub-key generated by the AES. Therefore, AESHA3 will be very relevant not only in terms of security but also it will save the resources in IoT devices. We implemented AESHA3 in intel core i7, 6th generation processor and Raspberry Pi 4B and find that up to 2 MB data encryption is very significant, and lesser the data size, more the resource saving compared to AES. In this we have made some initial analysis to compare the randomness (i.e. uniformity and independence) of the sub-keys generated by SHA-3 with AES generated sub-keys. The results are encouraging and the detail analysis will be published elsewhere.

\end{document}